\documentclass[aps,amsfonts,pra,twocolumn,showpacs]{revtex4}
  \usepackage{epsfig,amsmath,amssymb,bm,epsf,graphics,psfrag,verbatim}

\newcommand{\be}{\begin{equation}}
\newcommand{\ee}{\end{equation}}
\newcommand{\bea}{\begin{eqnarray}}
\newcommand{\eea}{\end{eqnarray}}

\newcommand{\lp}{\left(}
\newcommand{\rp}{\right)}

\def\prad{r_{0}}    
\def\lat{\ell}      

\begin{document}
\title{Phase-slip avalanches in the superflow
of $^4$He through arrays of nanopores}
\author{David Pekker}
\affiliation{Department of Physics, University of Illinois at Urbana-Champaign,
1110 West Green Street, Urbana, Illinois 61801-3080, USA}

\author{Roman Barankov}
\affiliation{Department of Physics, University of Illinois at Urbana-Champaign,
1110 West Green Street, Urbana, Illinois 61801-3080, USA}

\author{Paul M.~Goldbart}
\affiliation{Department of Physics, University of Illinois at Urbana-Champaign,
1110 West Green Street, Urbana, Illinois 61801-3080, USA}

\date{June 21, 2006}

\begin{abstract}
Recent experiments by Sato {\it et al.\/}~\cite{REF:Transition-2006}
have explored the dynamics of $^4$He superflow through an array of
nanopores.
These experiments have found that, as the temperature is lowered,
phase-slippage in the pores changes its character, from synchronous to
asynchronous.
Inspired by these experiments, we construct a model to address the
characteristics of phase-slippage in superflow through nanopore
arrays.
We focus on the low-temperature regime, in which the current-phase
relation for a single pore is linear, and thermal fluctuations may be
neglected.
Our model incorporates two basic ingredients: (1)~each pore has its
own random value of critical velocity (due, e.g., to atomic-scale
imperfections), and (2)~an effective inter-pore coupling, mediated
through the bulk superfluid.
The inter-pore coupling tends to cause neighbours of a pore that has
already phase-slipped also to phase-slip; this process may cascade,
creating an avalanche of synchronously slipping phases.
As the temperature is lowered, the distribution of critical velocities
is expected to effectively broaden, owing to the reduction in the
superfluid healing length, leading to a loss of synchronicity in
phase-slippage.
Furthermore, we find that competition between the strength of the
disorder in the critical velocities and the strength of the inter-pore
interaction leads to a phase transition between non-avalanching and
avalanching regimes of phase-slippage.
\end{abstract}
\maketitle

\noindent
Quantum phenomena at the macroscopic scale have been exhibited in a
variety of physical settings, including
superconductors~\cite{REF:Tinkham}, the helium
superfluids~\cite{REF:Tilley_Tilley}, and, more recently,
Bose-Einstein condensates (BEC) in atomic
vapours~\cite{REF:Dalfovo99}, and perhaps even
supersolids~\cite{REF:Kim_Chan}.  Spectacular examples include
persistent currents in multiply-connected superconductors and
superfluids, and the Josephson effects~\cite{REF:Josephson}, first
realised in superconducting tunnel junctions~\cite{REF:Josephson_exp2,
REF:Josephson_exp3} and, later, between two weakly coupled
BECs~\cite{REF:Albiez05}, as well as in superfluid
$^3$He~\cite{REF:Avenel88,REF:Backhaus_Packard, REF:Davis_Packard_RMP}
and $^4$He~\cite{REF:Sukhatme, REF:Hoskinson} weak links.

Widely-used descriptions of these systems are based upon the unifying concept
of spontaneous symmetry breaking~\cite{REF:Anderson_book}, implemented within
the framework of the Landau theory of phase transitions.  In this framework,
one introduces a suitable macroscopic Ginzburg-Landau order parameter
to characterise the broken symmetry state.  One major virtue of order-parameter
based descriptions is the view they afford of important topological processes,
such as vortex nucleation and growth, and the mechanisms for dissipation that
ensue.

Superconducting macroscopic quantum phenomena (MQP) have been employed
in many technologies, in fields ranging from biotechnology to radio
astronomy.  Primarily based on the DC Josephson effect---the essential
element in superconducting quantum interference devices
(SQUIDs)---these include SQUID amplifiers, metrology, biology and
medicine, nondestructive testing, geomagnetism, magnetic measurements
and microscopy, and prospecting for oil and
minerals~\cite{REF:Clarke-survey}.

The superfluid analogue of the SQUID differs in an essential way from
its superconducting counterpart: it is sensitive to interference
amongst (charge-neutral) atoms rather than (charged) electrons.  As a
result, it has considerable potential for use in measuring angular
velocities rather than magnetic fields and, therefore, for use as an
ultra-sensitive gyroscope~\cite{REF:Mukharsky, REF:Simmonds}.
As with SQUIDs, the essential element is the DC Josephson effect,
which has been realised in $^3$He~\cite{REF:Avenel88,REF:Backhaus_Packard} and,
more recently, in $^4$He~\cite{REF:Sukhatme, REF:Hoskinson}.  The latter is expected
to be more useful, technologically, owing to its relatively high
operating temperature, which exceeds that of $^3$He-based systems by a
factor of two thousand.

Recent experiments by the Berkeley group~\cite{REF:Transition-2006}
have explored a system that comprises two superfluid reservoirs
coupled via an array of pores, over a wide range of
temperatures~\cite{REF:WhyArrays}.  It was observed that, for
progressively lower temperatures below the critical temperature
$T_\lambda$ for the transition to the superfluid state,
the system passes through several regimes.
Just below \(T_\lambda\), for sufficiently small supercurrents, the array
was observed to function as a single effective Josephson junction
having an essentially sinusoidal current-phase relation.
For lower temperatures, below roughly a few millikelvin below
\(T_\lambda\), it was observed that the single-pore dynamics becomes
irreversible. This is due to the dissipative phase-slips that occur
whenever the superflow velocity in a pore reaches a critical
value~\cite{REF:Feynman,REF:Anderson66, REF:ILF1, REF:ILF2}, which we
believe to be specific to that pore.
At the high-temperature end of this regime, there is a narrow interval
of temperatures, of width roughly $10\,{\rm mK}$, within which all
pores appear to phase-slip simultaneously, and which Sato et
al.~\cite{REF:Transition-2006} refer to as the {\it synchronous regime\/}.
At lower temperatures, down to roughly $160\,{\rm mK}$ below
\(T_\lambda\), it appears that simultaneity in phase-slipping is lost;
Sato et al.~refer to this as the {\it asynchronous regime\/}.
It is believed that technologically viable devices can be designed to
function in both the Josephson (i.e.~reversible) and phase-slippage
(i.e.~irreversible) regimes, provided a large enough fraction of pores slip
sufficiently simultaneously to produce a measurable
``whistle'' at the Josephson frequency~\cite{REF:Whistle-2005}.  We
remark that, in the setting of multi-link superconducting devices, it
has been shown that the irreversible regime can be utilised for
magnetic-field and related phase-sensitive
measurements~\cite{REF:Hopkins-2005}.

\begin{figure}
\includegraphics[width=8cm]{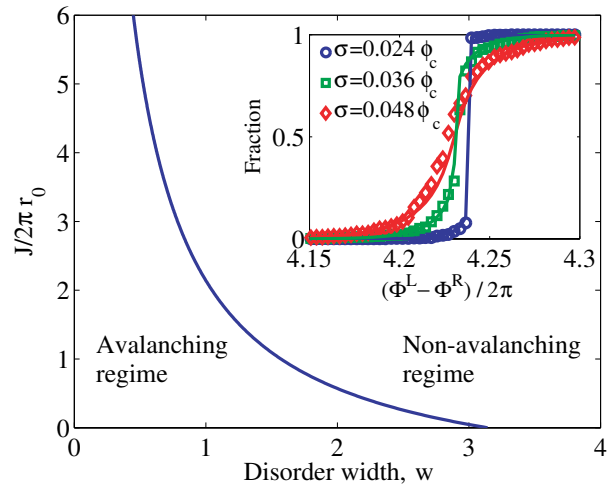}
\caption{Phase diagram showing avalanching and non-avalanching regimes
of the phase-slip dynamics, as a function of the effective pore
strength $J/2\pi r_0$ and disorder strength ${w}$. The diagram was
computed via our mean field theory, in the large-array limit, with
top-hat critical phase twist (or, equivalently, the velocity)
distribution of width \(w\).
{\it Inset:} The fraction of pores that have phase slipped, as a
function of the control-parameter $(\Phi^L-\Phi^R)/2\pi$. Comparison
between numerics for a \(25\times 25\) array of pores (points) and our
mean-field theory (lines) with Gaussian distribution of critical
velocities.}
\label{Fig:PD}
\end{figure}

In this Paper, we address the issue of the transition from synchronous
to asynchronous phase-slip dynamics of an array of nanopores
connecting a pair of superfluid reservoirs.  The main ingredients of
our description are pores that have random, temperature-dependent
critical velocities, along with an effective inter-pore coupling
mediated via superflow in the reservoirs.  We develop a model that
incorporates these ingredients, and analyse it via both a mean-field
approximation and exact numerical analysis for arrays consisting of a
relatively small number of pores.  Thus, we identify two effects:
(a)~strong disorder washes out the synchronicity of phase slips, which
leads to the loss of the ``whistle"; and (b)~if the disorder is sufficiently
weak, the phase-slip dynamics undergoes a disorder-driven phase
transition between avalanching and non-avalanching regimes
(see Fig.~\ref{Fig:PD}).  This enables us to obtain the current-phase
relation, as well as the amplitude of the current oscillations at the
Josephson frequency corresponding to a fixed chemical-potential
difference.  We believe that this model and our analysis of it
captures the essential physics taking place in the Berkeley group's
experiments~\cite{REF:Transition-2006}.

Disparate physical systems featuring competition between quenched
disorder and interactions, such as sliding tectonic
plates~\cite{REF:Burridge}, random-field
magnets~\cite{REF:RFIM}, and solids with disorder-pinned
charge-density waves (CDWs)~\cite{REF:Fisher98, REF:Marchetti}, are
well known to show phenomena analogous to the
avalanching-to-non-avalanching transition described here.  The model
and analysis that we employ are similar to those that have been
applied in the aforementioned settings~\cite{REF:kinship}.

\begin{figure}
\includegraphics[width=8cm]{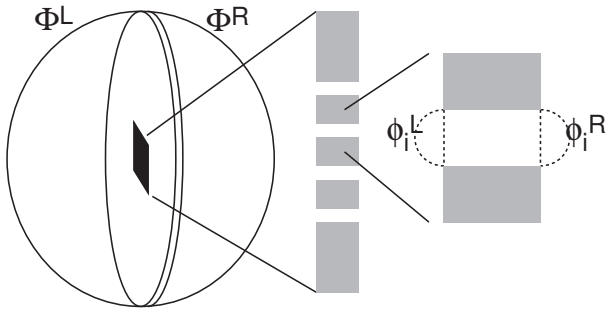}
\caption{Schematic of the model system. Left: the location of the pore
array on the membrane is indicated by the black region, and the phases
on hemispheres at infinity are labelled \(\Phi^\text{L}\) and
\(\Phi^\text{R}\). Centre: slice through the membrane, with pores
being represented by breaks in the membrane (white).  Right: boundary
conditions on hemispheres at the openings of the \(i\)-th pore.} \label{Fig:schematic}
\end{figure}

\noindent{\it Basic model\/}---%
The system we wish to describe consists of two reservoirs of
superfluid $^4$He, separated by a rigid barrier, embedded in which is
an array of pores, as shown in Fig.~\ref{Fig:schematic}. We shall
specialise to the case of an $N\times N$ array of pores, each having
radius $r_0$, situated at the sites of a square lattice of lattice
parameter $\ell$. It is straightforward to extend our analysis to
other array geometries.  It is convenient to regard the system as
comprising three components: two are bulk components [i.e.,~the left
(L) and right (R) reservoirs]; the third consists of the superfluid
inside the pores. We describe the state of the bulk helium in terms of
the superfluid order-parameter phase fields $\chi^{\text{L/R}}({\bf
r})$.  In doing this we are neglecting effects of amplitude
excitations of the order parameter, including vortices. In contrast,
within the pores we retain both amplitude and phase degrees of
freedom.  We imagine controlling the system by specifying the two
phases, $\Phi^{\text{L/R}}$, to which the phases in the left and right
reservoirs tend, far from the pore array. We believe that this level
of description allows us to capture the following important elements:
(a)~pores that exhibit narrow-wire-like meta-stable states, these
states being connected by phases slips; and (b)~interactions mediated
through the bulk superfluid in the two reservoirs, which couple pairs
of pores to one another and also couple the pores to the control
phases, $\Phi^{\text{L/R}}$.

To summarise, we describe the bulk superfluid helium reservoirs by the
phase-only Hamiltonians
\begin{equation}
H^{\text{L}/\text{R}}=
\frac{K_s}{2} \int_{\text{L/R}} d^3 r \,
\big\vert
\boldsymbol{\nabla}
\chi^{\text{L}/\text{R}}({\bf r})
\big\vert^2,
\label{Eq:HS}
\end{equation}
where \(K_s\equiv\hbar^2 n_s/m\) is the superfluid stiffness, in
which $n_s$ is the superfluid number density and $m$ is the mass of
a $^4$He atom.
To account for phase-slippage processes within a pore, which arise from
vortex lines crossing the pore, we shall use a modified phase-only
model that keeps track of the number of phase slips.  Therefore, we
take the energy of the superfluid inside the \(i^{\rm th}\) pore to be
\begin{align}
H_i=\frac{K_s}{2}J
\left(
\phi^\text{L}_i-\phi^\text{R}_i -2 \pi n_i
\right)^2,
\end{align}
in which \(J\equiv\pi\, r_0^2/d\) accounts for the geometry of the
pore, where \(d\) is of the order of the membrane thickness, and
\(\phi^{\text{L}/\text{R}}_i\) is the phase of the order parameter in
the vicinity of the left/right side of the \(i^{\rm th}\) pore.
Furthermore, \(n_i\) counts the net number of phase slips that would
occur in the \(i^{\rm th}\) pore if the system were to progress from
a reference state in which the phase were uniform throughout.
This description of the pores must be supplemented by specifying the
critical velocities at which phase slips occur, as we shall discuss in
detail below.

To reduce the model to one involving only the phase-differences across
the pores and the phases at infinity, we minimise the
energy~(\ref{Eq:HS}) in the reservoirs, which forces
$\chi^{\text{L}/\text{R}}({\bf r})$ to obey the Laplace equation.
Focusing on the left reservoir, the appropriate boundary conditions
are (see Fig.~\ref{Fig:schematic}): (a)~the phase on the hemisphere at
infinity is \(\Phi^{\text{L}}\); (b)~no current flows through the
membrane surface between the pores,
i.e.~\(\nabla_\perp\chi^\text{L}=0\) there; and (c)~the phase on the
hemisphere of radius $r_0$ centred at the opening of the \(i^{\rm
th}\) pore is specified to be \(\phi_i^\text{L}\).  The choice of
surface for the last of the boundary conditions is made for
convenience, as it simplifies the resulting mixed boundary value
problem whilst enforcing the physical condition of continuity of the
phase in the vicinity of the pore opening.

To solve this mixed boundary value problem we appeal to its
electrostatics analogy, in which the phase and the superfluid
stiffness $K_s$ respectively play the roles of the scalar potential
and the permittivity $\epsilon_0$.  To compute the energy at the
minimum in terms of the boundary data we apply the divergence theorem
to Eq.~(\ref{Eq:HS}) to obtain the energy in terms of a surface
integral:
\begin{equation}
H^{\text{L}}=\frac{K_s}{2} \int_{\partial\, \text{L}}
d{\bf S}\cdot
\chi^{\text{L}} \boldsymbol{\nabla}\chi^{\text{L}}-
\frac{K_s}{2}\int_{\text{L}} d^3r\,\chi^{\text{L}}\,\nabla^2 \chi^{\text{L}},
\end{equation}
where the volume (i.e.~last) term vanishes because at the
minimum-energy configuration $\chi^\text{L}$ obeys
\(\nabla^{2}\chi^\text{L}=0\).  To simplify the evaluation of the
surface integral we make use of the fact that the energy is invariant
under global shifts of the potential.  Thus, we lower all potentials
by $\Phi^{\text{L}}$, which eliminates the contribution from the (left)
surface at infinity.  What remains are the contributions from the
membrane: those from between the pores give zero, owing to the
zero-flux boundary condition; those from those from the hemispherical
surfaces covering the pores give
\begin{equation}
H^{\text{L}}=\frac{1}{4} \sum_i (\phi_i^\text{L}-\Phi^\text{L}) \, q_i,
\label{Eq:Uc}
\end{equation}
where, by assumption, the phase $\chi^\text{L}$ takes the constant
values $\phi_i^\text{L}$ on the hemispherical surfaces, and the
remaining surface integrals have been replaced (via the analogue of
Gauss's law) by half of the (analogue of) the charge enclosed.  For
pore $i$ this charge is $q_i\equiv K_s\int d{\bf
S}\cdot\boldsymbol{\nabla}\chi^{\text{L}}$, where the integral is
taken over the spherical surface that completes the hemispherical one.
Reflecting the problem across the membrane's left surface, we see that
we are looking for half of the electrostatic energy of a set of
suitably charged metal spheres at potentials
\(\phi_i^\text{L}-\Phi^\text{L}\), and what remains is to determine
the charges \(q_i\).  To do this, we consider the \(i^{\rm th}\)
sphere: in the limit $\lat\gg\prad$ (i.e.~ignoring di- and
higher-order charge-multipoles), the potential on its surface obeys
\begin{subequations}
\begin{align}
&K_s \left(\phi_i^\text{L}-\Phi^\text{L}\right)=
\sum_j C^{-1}_{ij} q_j,
\label{Eq:imp-charge}
\\
&C^{-1}_{ij}
\equiv
\frac{\delta_{ij}}
{4 \pi \, r_0}+\frac{1-\delta_{ij}}{4 \pi \,|r_{ij}|},
\label{Eq:C-def}
\end{align}
\end{subequations}
where \(|r_{ij}|\) is the distance between the \(i^{\rm th}\) and
\(j^{\rm th}\) pore (sphere) centres, and \(C_{ij}\) is the analogue of the
{\it capacitance matrix\/}.
By solving Eq.~(\ref{Eq:imp-charge}) for the \(q_i\)'s and eliminating
them from Eq.~(\ref{Eq:Uc}), we arrive at the combined energy of the
left and right superfluid reservoirs and the superfluid in the pores:
\begin{align}
E=\frac{K_s}{2} \sum_{i j} (\phi_i^\text{L}-\Phi^\text{L})\, C_{i j}\,
(\phi_j^\text{L}-\Phi^\text{L}) + \sum_i H_i\,,
\label{Eq:totalEnergy}
\end{align}
where we have, without loss of generality, restricted our attention to
spatially anti-symmetric states, for which
\(\Phi^\text{R}=-\Phi^\text{L}\) and
\(\phi^\text{R}_i=-\phi^\text{L}_i\).

We complete the description of our model by specifying the single-pore
dynamics, and thus the mechanism by which energy is dissipated in the
pores.  The superfluid velocity \(v_i\) in a pore of thickness $d$ is
defined by the phases at the pore openings: $v_i=\hbar\nabla\phi_i/m
\approx \hbar\lp 2 \phi^{\text L}_{i}-2\pi n_{i}\rp/dm$.
Correspondingly the current through the pore is given by
\begin{equation}
I_i=\frac{K_s J}{\hbar}\lp 2 \phi^{\text
L}_{i}-2\pi n_{i}\rp.
\end{equation}
When the velocity through the $i^{\rm th}$ pore exceeds its critical
value $v_{c,i}$ (or, equivalently, \(\phi^\text{L}_i-\pi n_i\) exceeds
\(\phi_{c,i}\)), a vortex line nucleates and moves across the
pore, which decreases the phase-difference across the pore by \(2 \pi\).
Thus, immediately after a phase slip, the velocity through the \(i^{\rm
th}\) pore is decreased by $\Delta v\equiv 2\pi\hbar/md$, a quantum of
superfluid velocity drop for a pore having fixed phases at the openings,
i.e.~$v_i\to v_i-\Delta v$. However, after a very short time,
controlled by the speed of sound, the system balances the superflow in
the bulk reservoirs and through the various pores.  Therefore, after
this relaxation process is complete, the actual drop in the superfluid
velocity through the \(i^{\rm th}\) pore is always less than $\Delta
v$.  To find the configuration of the superfluid after a phase slip,
we note that the phase-difference along a path from the left
hemispherical surface at infinity through the $i^{\rm th}$ pore to the
right hemispherical surface at infinity drops by \(2 \pi\), whilst the
phase-difference along a path through any other pore remains
unaffected.  In our model, we implement this kind of phase-slip event
by sending $n_{i}$ to $n_{i}+1$ (assuming all flow is to the left) and
finding a new set of values for all of the \(\phi^\text{L}_i\)'s by
minimising the total energy, Eq.~(\ref{Eq:totalEnergy}).  

Not too near $T_\lambda$ the experimentally-observed
temperature-dependence of the critical velocity for superflow through
a pore is approximately linear~\cite{REF:Beecken_Zimmermann,
REF:Avenel93}: $v_c(T)=v_c(0)(1-T/T_0)$, where $T_0\approx
2.5{\rm\,K}$.  This $T$-dependence is most closely reproduced by the
thermal nucleation of half-ring vortices mechanism~\cite{REF:Volovik}.
However, the linear $T$-dependence does not hold in the temperature regime that
we are primarily interested in; instead, the critical velocity is
proportional to the superfluid stiffness there~\cite{REF:Hess71,REF:ILF1,
REF:ILF2}: 
\be\label{Eq:V_c_est} v_c(T)\simeq\hbar/m\xi(T)\propto
K_s(T), 
\ee 
where $\xi(T)\simeq\xi_{0}(1-T/T_{\lambda})^{-2/3}$ is the
temperature-dependent healing length.

Extrinsic effects are known to reduce critical velocities from the
intrinsic values discussed above.  We hypothesise that in the Berkeley
group's experiments the extrinsic effects originate in atomic-scale roughness
of the pore walls, and play a pivotal role in generating
critical-velocity variability amongst the pores.  This variability is
expected to be temperature dependent because only roughness on
length-scales longer than $\xi(T)$ can substantially perturb the order
parameter and thus influence the critical velocities.  Hence, at
higher temperatures the impact of surface roughness is expected to be
weaker and, correspondingly, the distribution of critical velocities
is expected to be narrower.  Thus, lowering the temperature has the
important effective consequence of increasing the effective
disorder~\cite{REF:SP-pc}.

\begin{figure}
\includegraphics[width=8cm]{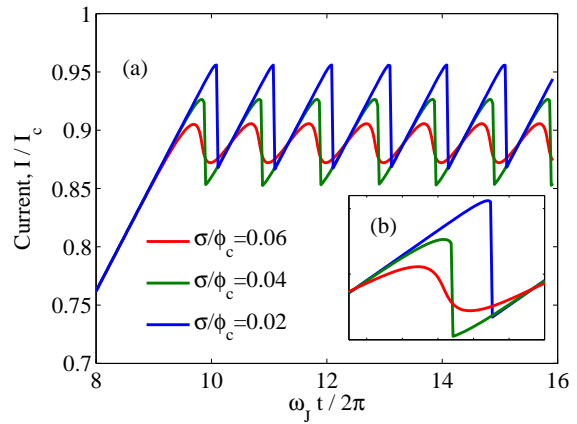}
\caption{Traces of total current through an array of pores as a
function of time, at various disorder strengths~\cite{REF:parameters}.
The traces were computed at fixed chemical potential difference
\(\Delta \mu\), via our mean-field theory, with Gaussian distributions
of critical phase-twists \(\phi_{c,i}\) of widths \(\sigma\) and mean
\(\phi_c=3\pi\).  The corresponding critical current, in the absence
of disorder, is given by \(I_c=2 K_s J \phi_c/\hbar\). As the disorder
strength is increased, the amplitude of current oscillations
decreases. The sharp drops in the current, which correspond to
avalanches, disappear for \(\sigma/\phi_c>0.052\).}
\label{Fig:current}
\end{figure}

\noindent{\it Implications of the model \/}---%
We shall work at fixed (negative) difference \(\Delta \mu\) in the
chemical potential between the reservoirs, so that the control
parameter \(\Phi^\text{L}\) evolves linearly in time, according to the
Josephson-Anderson relation
\begin{align}
\Phi^\text{L}=-\Phi^\text{R}=-\frac{\Delta \mu}{2 \hbar} \, t.
\label{Eq:Josephson}
\end{align}
As \(\Phi^\text{L}-\Phi^\text{R}\) grows, so do the superfluid
velocities through the various pores, punctuated at regular intervals
by velocity drops associated with the phase-slip processes.
As the total energy of the state is periodic in \(\Phi^L\) with period
\(\pi\), the total current through the array must be a periodic
function of time with the period given by the Josephson frequency
\(\omega_\text{J}=\Delta \mu/\hbar\).  Due to the randomness of the
critical velocities amongst the pores, the velocities in the various
pores do not reach their critical values simultaneously.  Therefore,
the pores having the smaller critical velocities (i.e.~the weaker
pores) tend to slip first.  If the distribution of critical velocities
is sufficiently narrow, the array may, as we demonstrate below, suffer
an avalanche. By an avalanche we mean that when the weaker pores slip,
superflow through the neighbouring pores that have yet to slip
increases, due to the inter-pore interaction, and this drives them to
their own \(v_{c,i}\), causing a cascade of phase slips in which an
appreciable fraction of pores in the array slip.  Experimentally,
avalanches are reflected in a periodic series of sharp drops in the
total current through the array of pores as a function of time.
Time-traces of the total current in the avalanching and
non-avalanching regimes are contrasted in Fig.~\ref{Fig:current}.

For arrays having a small number of pores, the quasi-static state of
(mechanical) equilibrium may be numerically determined, as the control
parameter $\Phi^{\text{L}}-\Phi^{\text{R}}$ evolves parametrically.
As a consequence of the long-range nature of the inter-pore couplings,
the array dynamics is well approximated by a mean-field theory.  We
shall describe a numerical approach first, and then the mean-field
theory.

The numerics take as input: $J$, the $v_{c,i}$'s, and the
(non-inverted) matrix \(C^{-1}_{ij}\) [see Eq.~(\ref{Eq:C-def})],
which itself depends on \(r_0\), \(l\), and \(N\).  At each time-step,
\(\Phi^\text{L}\) is incremented, and the new \(\phi_i^\text{L}\)'s
are obtained.  If the superflow through any pore is found to now
exceed its critical velocity, that pore phase-slips (i.e.~its value of
\(n_i\) is incremented by plus unity); next, the various
\(\phi_i^\text{L}\)'s are recomputed using the new set of \(n_i\)'s,
and the program goes back to recheck if any other pore now exceeds its
critical velocity. This continues until no new phase-slips are found
to occur, at which point the control parameter is incremented and the
procedure is repeated.

\begin{figure}
\vspace{2mm}
\includegraphics[width=8cm]{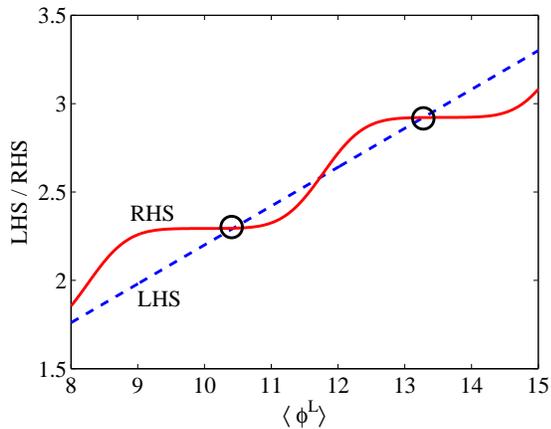}
\caption{Left- and right-hand sides of the averaged form of
Eq.~(\ref{Eq:phii}).  
The circles indicate the two main solutions: as the control-phase is
increased, the RHS curve moves up and to the left, such that there is
a discontinuous jump from one solution to the other corresponding to
an avalanche.  }
\label{Fig:graphicalSolution}
\end{figure}

Next, we construct a mean-field theory.  We assert that the
control phase-difference is monotonically increasing in time, so that
phase differences \(\phi^\text{L}_i-\phi^\text{R}_i\) are always
increasing and the superflow in the pores undergoes only
\(n_i\)-increasing phase slips.  Then we may proceed by selecting an
arbitrary pore $i$, and minimising the energy~(\ref{Eq:totalEnergy})
with respect to $\phi^{\text{L}}_{i}$; by replacing the
$\phi^{\text{L}}_{j(\ne i)}$'s by the mean-field value
$\langle\phi^{\text{L}}\rangle$ we obtain an equation for the phase
at the \(i^{\rm th}\) pore:
\begin{align}
\phi_i^\text{L}(C+2 J)- B \langle \phi^\text{L} \rangle
={A \Phi^\text{L} +2 \pi J \, n_i},
\label{Eq:phii}
\end{align}
where \(A\equiv\sum_i C_{ij}\), \(B\equiv C_{ii}-A\), and \(C\equiv
C_{ii}\)~\cite{REF:exp-TI}.  (The necessary inversion of $C_{ij}^{-1}$
can readily be accomplished either analytically, by transforming to
Fourier space, or numerically.)\thinspace\ By averaging the left- and
right-hand sides of this equation over sites, we arrive at a
self-consistency condition on \(\langle \phi^\text{L} \rangle\).
Next, by assuming self-averaging with respect to the disorder in the
critical velocities, we may replace the average over sites by an
average over disorder.  This procedure amounts to replacing, in the
above condition, \(\phi^\text{L}_i\) by \(\langle\phi^\text{L}
\rangle\) and \(n_i\) by
\begin{widetext}
\begin{equation}
\langle n \rangle (\langle \phi^\text{L} \rangle) =
\sum_{k} k
\int_{0}^{\infty} d\phi_c\,Q(\phi_c)
\left[
\Theta \left (\phi_c-\frac{A \Phi^\text{L} + B \langle \phi^\text{L} \rangle-\pi k \,
C}{{C+2 J}}\right)-
\Theta \left( \phi_c-{\frac{A \Phi^\text{L} + B
\langle \phi^\text{L} \rangle-\pi (k-1) C}{C+2
J}}\right) \right],
\end{equation}
\end{widetext}
where \(Q(\phi_c)\) is the probability distribution of half-critical phase
twists in the pores \(\phi_c \equiv d m v_c/\hbar\).  The averaged form of
Eq.~\ref{Eq:phii} can be solved graphically, by plotting the left- and
right-hand sides as functions of \(\langle \phi^\text{L}\rangle\); see
Fig.~\ref{Fig:graphicalSolution}.
It is evident from this graphical approach that whenever the maximum
slope of the right-hand side, \(2 \pi J \partial \langle n \rangle /
\partial \langle \phi^\text{L} \rangle\) {\it fails\/} to exceed the
slope of the left-hand side, \((A+2J) \langle \phi^\text{L} \rangle
\), the self-consistency condition yields a unique solution for the
average phase $\langle \phi^\text{L} \rangle$, and that this phase
evolves continuously with the (increasing) control phase
$\Phi^\text{L}$.  This corresponds to the non-avalanching regime.
By contrast, whenever the maximum slope of the right-hand side {\it
does\/} exceed the slope of the left-hand side, the self-consistency
condition no longer yields a unique solution for $\langle
\phi^\text{L} \rangle$.  Instead, as the control-phase increases, the
continuous evolution of $\langle \phi^\text{L} \rangle$ is punctuated
by jumps, which occur when pairs of solutions merge and disappear.
These jumps reflect avalanching behaviour, and we refer to this as the
avalanching regime.

One can use this mean-field theory to construct a phase diagram that
demarcates avalanching and non-avalanching regimes, for any choice of
disorder distribution $Q$.  For the case of a top-hat
distribution of critical phase-twists, a simple inequality defines the
avalanching regime:
\begin{equation}
w\le w_\text{c}\equiv\frac{2 \pi J B}{(A + 2 J)(C + 2 J)},
\end{equation}
where \(w\) is the width of the top hat, and \(w_\text{c}\) is its
critical value; such a phase boundary is exemplified in
Fig.~\ref{Fig:PD}.  In experiments, one can explore this phase diagram
by tuning the temperature, which, as we have hypothesised above,
effectively tunes the strength of the disorder. 

Thermal fluctuations of the phases wash out the disorder-driven phase
transition if they exceed the width of the disorder distribution.
To avoid this, the temperature has to be smaller than the energy cost
of winding the phase of a single pore by the critical disorder width,
i.e.~\(k_\text{B} T\lesssim K_s(T)\,C \, w_\text{c}^2/2\). We estimate
that this inequality is satisfied provided that $T$ is not too close
to the $\lambda$-point, i.e.~\(T_\lambda-T\gtrsim 3\,\text{mK}\) in the
setting of Ref.~\cite{REF:Transition-2006}. Furthermore, we have
assumed that the current-phase relation is linear, which is true
provided that \(T_\lambda-T \gtrsim 5 \, \text{mK}\), as was measured
for a similar setup in Ref.~\cite{REF:Hoskinson}.

To test the results of the mean-field theory, we have compared them to
results from a numerical investigation performed on a finite, periodic
lattice~\cite{REF:exp-TI}.  For various widths of the disorder
distribution (which we have taken to be Gaussian), the fractions of
pores that have phase-slipped, as a function of control-parameter, are
shown in the inset of Fig.~\ref{Fig:PD}.  As predicted by the
mean-field theory, avalanches occur when the distribution of critical
velocities is narrow but not when it is broad.  At a critical strength
of the disorder, which separates these two regimes, the discontinuity
in the mean-field fraction of slipped pores just vanishes.  Moreover,
at this critical disorder there are always values of the control phase
at which the response of the system diverges.  The mean-field and
numerical results appear to agree with one another rather well, as one
can see from Fig.~\ref{Fig:PD}, at least in the vicinity of the
critical point.

\begin{figure}
\includegraphics[width=8cm]{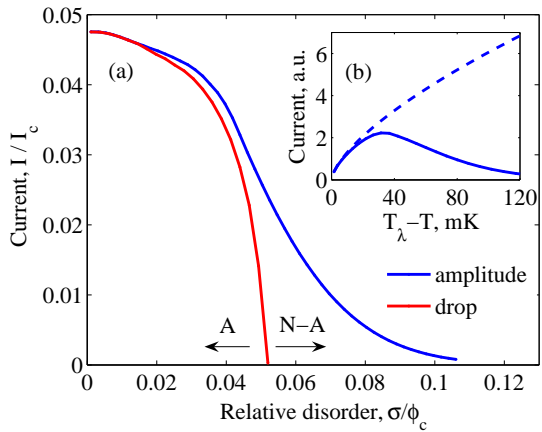}
\caption{Amplitude of the oscillations of the current \(\Delta I\), and
drop in the current caused by an avalanche, as functions of the disorder
strength. A and N-A indicate the avalanching and non-avalanching
regimes, respectively.  
{\it Inset:\/} Solid line: amplitude of the current oscillations as a
function of temperature, using the disorder model described in the
text.  Dashed line: expected amplitude in the absence of disorder.  }
\label{Fig:order_synch}
\end{figure}

The two main results of our Paper are summarised in
Fig.~\ref{Fig:order_synch}.  The blue curve shows the dependence of
the amplitude of the current oscillation \(\Delta I\) (i.e.~half the
distance between smallest and largest current during a single period
in Fig.~\ref{Fig:current}) on the disorder strength.  As the disorder
becomes stronger, the phase-slips in the various pores become more
asynchronous, and the oscillations in the current gradually disappear.
The red curve shows the dependence of the current drop caused by an
avalanche (i.e.~the height of the vertical drop in current in
Fig.~\ref{Fig:current}) on the disorder strength.  The current-drop
plays the role of an order-parameter in a second-order phase
transition tuned by the disorder strength.  As the disorder becomes
stronger, the order-parameter decreases, becoming zero at the critical
disorder strength corresponding to the transition from the avalanching
to the non-avalanching regime (i.e.~\(\sigma_c\approx 0.052 \phi_c\)).

\noindent{\it Comparison with experiments\/}---%
In their experiments~\cite{REF:Transition-2006}, the Berkeley group
have measured the amplitude of the ``whistle'' (i.e.~the amplitude of the
current oscillations \(\Delta I\)) as a function of temperature at a
fixed chemical potential difference.  The experiments found that at
\(T_\lambda\) there are no oscillations of the current.  As the
temperature is lowered below \(T_\lambda\), \(\Delta I\) first
increases, and then gradually decreases.  To obtain the behaviour of
\(\Delta I\) as a function of \(T\) we extend the model for the critical
velocity, Eq.~(\ref{Eq:V_c_est}), to include disorder, by assuming that
\begin{equation}
v_{c,i}(T)\simeq\frac{\hbar}{m(\xi(T)+x_i)},
\end{equation}
where \(x_i\) is the temperature-independent characteristic scale of
the surface roughness in the \(i^{\rm th}\) pore, which we take to
have a Gaussian distribution~\cite{REF:distribution}.  For
\(T_\lambda-T>5\,\text{mK}\), we can compare the results of our
modelling to those of the experiments.  The general features are
reproduced: the initial increase in \(\Delta I\) is associated with an
increase in the superfluid fraction; the gradual decrease at lower
temperatures is due to the loss of synchronicity amongst the pores,
which is caused by the effective increase in the strength of disorder.
To demonstrate these features, we plot the amplitude of current
oscillations as a function of temperature; see the inset in
Fig.~\ref{Fig:order_synch}, for which we have set \(\delta
x=4\,\text{nm}\).

We also note that the general features of the current-{\it vs.\/}-time
traces, Fig.~\ref{Fig:current}, are similar to those of the type III
experiments described in Ref.~\cite{REF:Transition-2006}. In both, as
the temperature is lowered (i.e.~the disorder is increased), the
avalanche gradually disappears, and then so do the oscillations in the
current.

\noindent{\it Concluding remarks\/}---%
Motivated by recent experiments performed by the Berkeley group on
superflow through nanopore arrays~\cite{REF:Transition-2006}, we have
developed a model to describe phase-slip dynamics of such systems.
The main features of our model are effective inter-pore couplings,
mediated through the bulk superfluid, as well as randomness in the
critical velocities of the pores, the latter being effectively
controlled through the temperature.

Within our model, we find that the competition between
(a)~site-disorder in the critical velocities and (b)~effective
inter-pore coupling leads to the emergence of rich collective
dynamics, including a transition between avalanching and
non-avalanching regimes of the phase-slip dynamics.  We identify a
line of critical disorder-strengths in the phase diagram, at which
there is a divergent susceptibility, in the sense that near to this
line small changes in the control parameter can lead to large changes
in the fraction of phase-slipped pores.

Our model reproduces the key physical features of the Berkeley group's
experiments~\cite{REF:Transition-2006}, including a high-temperature
synchronous regime, a low-temperature asynchronous regime, and a
transition between the two.  We therefore feel that the model captures
the essential physics explored in these experiments.

In a forthcoming paper we shall extend our approach in order to
address the transition from the Josephson regime to the avalanching
phase-slippage regime described here, by including thermal
fluctuations together with a generalised description of pore energies
that is valid in both regimes.

\smallskip
\noindent
{\it Acknowledgements\/}---%
It is a pleasure for us to express our gratitude to Richard Packard
for introducing us to the subject-matter explored in this Paper, for
describing his group's experiments to us, and for furnishing us with
data of theirs, prior to publication.  We are similarly grateful to
his group members Yuki Sato and Aditya Joshi.
We thank Karin Dahmen for guiding us through the literature on
analogous physical systems, and Tzu-Chieh Wei and Smitha Vishveshwara
for useful discussions.
This work was supported by the U.S.~Department of Energy, Division of
Materials Sciences under Award No.~DEFG02-96ER45434, through the
Frederick Seitz Materials Research Laboratory at the University of
Illinois at Urbana-Champaign.


\begin{thebibliography}{99}
\bibitem{REF:WhyArrays} For $^4$He, thermal fluctuations strongly
suppress the DC Josephson effect in a single
pore~\cite{REF:Zimmermann, REF:Chui}.  To overcome this difficulty, experiments
are conducted using arrays of pores; see, e.g.,
Ref.~\cite{REF:Sukhatme}.
\bibitem{REF:kinship} In particular, one can make evident the kinship
between the present superfluid model and the elastic model for CDW
depinning (without topological excitations) by exchanging control via
the chemical-potential difference between the reservoirs for control
via the supercurrent, which then plays the role of the CDW driving
force.
\bibitem{REF:exp-TI} We wish to eliminate effects arising from edges
of the array of pores.  We do this by measuring distances between
pairs of sites, allowing paths to wrap around the edges and picking
the shortest one.  This renders the problem translationally invariant.
In particular, $A$, $B$, and $C_{ii}$ are then independent of any pore
index.
\bibitem{REF:distribution} The Gaussian distribution is suitably
cut-off and renormalized to avoid negative or very large critical
velocities.
\end{thebibliography}
\end{document}